\documentclass[aps,pre,twocolumn]{revtex4-1}
\usepackage[dvips]{graphicx}
\usepackage{amsmath}
\usepackage{verbatim}
\usepackage{color}
 \graphicspath{{./Figures/}}
\newcommand{\br}{{\bf r}}

\newcommand{\la}{\left\langle}
\newcommand{\ra}{\right\rangle}

\begin{document}

\title{Osmotic Swelling Behavior of Surface-Charged Ionic Microgels}

\author{Mohammed O. Alziyadi}
\altaffiliation[Current address: ]{Department of Physics, College of Science and Humanities-Shaqra, Shaqra University, Riyadh, Saudi Arabia}
\affiliation{Department of Physics, North Dakota State University, Fargo, ND 58108-6050, USA}
\author{Alan R. Denton}
\email[]{alan.denton@ndsu.edu}
\affiliation{Department of Physics, North Dakota State University, Fargo, ND 58108-6050, USA}

\begin{abstract}
In recent years, ionic microgels have garnered much attention due to their unique properties, 
especially their stimulus-sensitive swelling behavior. The tunable response of these soft,
permeable, compressible, charged colloidal particles is increasingly attractive for applications 
in medicine and biotechnologies, such as controlled drug delivery, tissue engineering, and 
biosensing. The ability to model and predict variation of the osmotic pressure of a single 
microgel with respect to changes in particle properties and environmental conditions proves vital
to such applications. In this work, we apply both nonlinear Poisson-Boltzmann theory and 
molecular dynamics simulation to ionic microgels (macroions) in the cell model to compute 
density profiles of microions (counterions, coions), single-microgel
osmotic pressure, and equilibrium swelling ratios of spherical microgels whose fixed charge 
is confined to the macroion surface. The basis of our approach is an exact theorem that 
relates the electrostatic component of the osmotic pressure to the microion density profiles.
Close agreement between theory and simulation serves as a consistency check to validate our approach. 
We predict that surface-charged microgels progressively deswell with increasing microgel
concentration, starting well below close packing, and with increasing salt concentration,
in qualitative agreement with experiments. Comparison with previous results for microgels 
with fixed charge uniformly distributed over their volume demonstrates that surface-charged 
microgels deswell more rapidly than volume-charged microgels. We conclude that swelling 
behavior of ionic microgels in solution is sensitive to the distribution of fixed charge 
within the polymer-network gel and strongly depends on bulk concentrations of both 
microgels and salt ions. 
\end{abstract}
 
\maketitle

\section{Introduction}
Microgels are soft colloidal particles, made of a cross-linked polymer network gel, that become 
swollen in a good solvent~\cite{MicrogelBook2011,HydrogelBook2012,karg-richtering2019,scotti2022}.
Depending on the procedure for their chemical synthesis~\cite{pelton2011}, microgels can be 
designed to swell/deswell in response to external parameters, such as temperature, $p$H, and 
ionic strength~\cite{Fernandez-Barberoch4}. Swelling behavior depends on mechanical (elastic) 
properties of the gel and polymer-solvent interactions~\cite{P.J.Flory,weitz2011}. 
When microgels are dispersed in a polar solvent (e.g., water), dissociation of counterions 
from the polymer chains can turn the particles into colloidal macroions with a fixed 
(immobile) charge~\cite{nieves-jcp2003,likos2011}. Swelling behavior of ionic microgels
is influenced by electrostatic interactions between fixed charge, counterions, and coions 
(from added salt)~\cite{weitz-jcp2012}. Microgels and, in particular, ionic microgels have 
been extensively investigated due to their applications in biomedical, food, pharmaceutical, 
and petroleum industries~\cite{pashkovski2011,malmsten2011,ben-wang2011,tavakoli2017hydrogel,
hamidi2008hydrogel}. 

The swelling behavior of microgels, in contrast to that of bulk 
gels~\cite{tanaka1982,tanaka1984,siegel1988,barrat-joanny-pincus1992},
is fundamentally determined by the single-particle osmotic pressure, defined as the difference 
in pressure between the interior and exterior regions~\cite{denton-tang2016,denton-alziyadi2019}. 
In thermodynamic equilibrium, this single-particle osmotic pressure must vanish, achieved only 
through a delicate balance between electrostatic, elastic, and mixing entropic contributions.
In concentrated suspensions of microgels, equilibrium particle sizes respond also to 
self-crowding, as each particle adjusts its internal degrees of freedom to the local 
environment, with relevance for macroscopic phase behavior.

Swelling properties of microgels (ionic and nonionic) have been investigated by several
complementary experimental methods, including light scattering~\cite{nieves-macromol2000,
mohanty-richtering2008,holmqvist-shurtenberger2012,braibanti-perez2016}, 
neutron scattering~\cite{stieger2004,nieves-jcp2010,Nojd2018,zhou2023}, 
osmometry~\cite{nieves-prl2015,scotti2021}, and dielectric spectroscopy~\cite{dhont2016}. 
Compressible microgels and suspensions
thereof have been modeled by a variety of theoretical methods, including mean-field 
theories~\cite{cloitre-leibler1999,cloitre-leibler2003,levin2002,schurtenberger-ZPC2012,
colla-likos2014}, integral-equation theories~\cite{moncho-jorda2013}, effective interaction 
theories~\cite{denton2003,gottwald2005,weyer-denton2018,aguirre-manzo-perez2019}, and 
cell model-based theories~\cite{denton-tang2016,denton-alziyadi2019,brito2019}. 
Simulation studies have applied various molecular 
dynamics~\cite{hedrick-chung-denton2015,winkler2014,winkler2017,schurtenberger2014,
zaccarelli2017} and Monte Carlo~\cite{schneider-linse2002,molina2013,schneider2017,
urich-denton2016,weyer-denton2018} methods. From past work, swelling behavior is known 
to depend sensitively on single-particle properties, including the distribution of cross-links 
in the polymer network comprising the gel~\cite{schurtenberger-SM2012,boon-schurtenberger2017}. 
In general, the lower the cross-link concentration, the softer and more compressible the particles,
although the spatial distribution of cross-links is also relevant.

Less well understood is the connection between swelling and the distribution of fixed charge
on the polymer network. In chemical synthesis protocols developed to produce ionic microgels,
the fixed charge distribution can be controlled, to some extent, by adjusting the concentration 
and chemical species of the initiator for the polymerization reaction in the synthesis. 
In some microgel systems, initiator molecules themselves are charged and tend to 
deposit near the particle's periphery, yielding a surface-localized fixed 
charge~\cite{nieves-pnas2016,scotti-binary-pre2017,gasser2019,scotti2021,Zhou2012}. 
How in detail the fixed charge distribution affects osmotic pressure and equilibrium swelling 
of ionic microgels is a largely unresolved question.

To address the question of how fixed charge distribution influences swelling of microgels, 
we have developed theoretical and computational modeling methods for ionic microgels 
and applied them to macroions with fixed charge confined to their surface. 
Comparison of predictions for this idealized model with corresponding
results from our earlier studies of microgels with fixed charged uniformly distributed
throughout their volume~\cite{denton-tang2016,denton-alziyadi2019} contrasts the 
sensitivity of swelling to fixed charge distribution between two extreme distributions.
Our approach is based on fundamental statistical mechanical relations for the electrostatic 
component of the single-microgel osmotic pressure that accurately account for counterion and 
coion correlations.  Building on our previous work, this paper describes the swelling of ionic 
microgels as a competition between electrostatic and gel contributions to the osmotic pressure 
and demonstrates the sensitivity of swelling to the fixed charge distribution.

In Sec.~\ref{models}, we define the primitive and cell models of ionic microgels, which are
the foundation for our study. In Sec.~\ref{theory}, we review our theory of microgel swelling, 
deriving the electrostatic component of the single-microgel osmotic pressure from a 
partition function and approximating the gel contribution using the Flory-Rehner theory 
of polymer networks. Section~\ref{methods} outlines our computational implementation
of the swelling theory via Poisson-Boltzmann theory and molecular dynamics simulation. 
Section~\ref{results} presents numerical results for counterion density profiles, 
single-microgel osmotic pressure, and equilibrium swelling ratios of surface-charged microgels. 
We compare and contrast the swelling behavior with that of volume-charged microgels 
and discuss the influence of added salt. Finally, Sec.~\ref{conclusions} summarizes 
our results and concludes with suggestions for future work.

\section{Models}\label{models}
\subsection{Primitive Model}
\begin{figure}[t]
\includegraphics[width=0.8\columnwidth,angle=0]{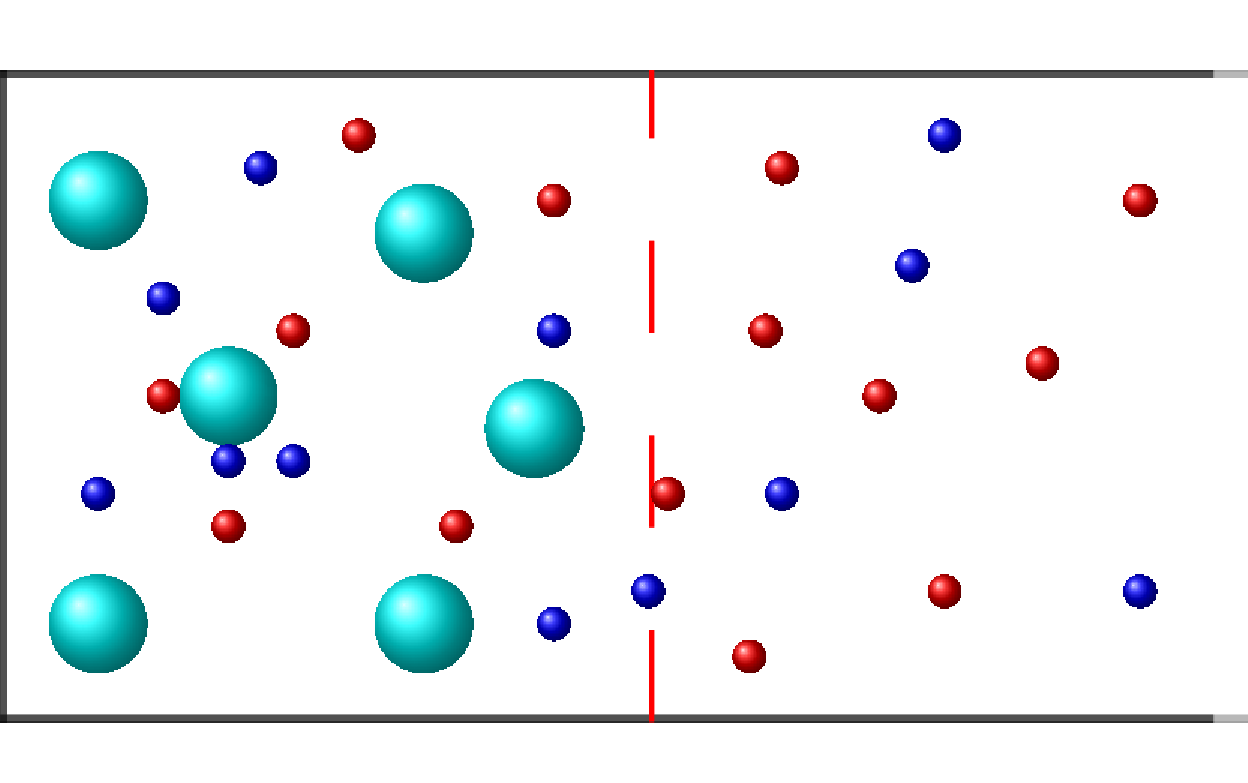}
\includegraphics[width=0.8\columnwidth,angle=0]{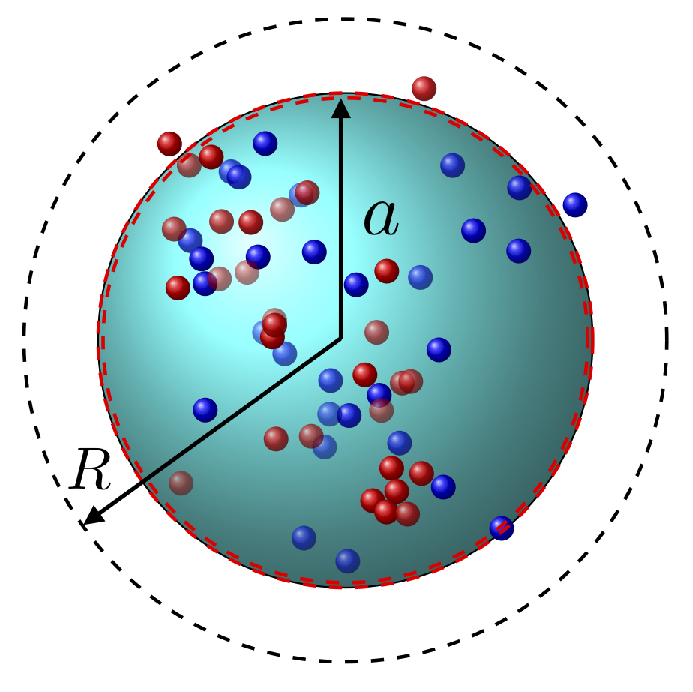}
\vspace*{-0.2cm}
\caption{Schematic drawings of (top) primitive model of a suspension of spherical ionic microgels 
(large cyan spheres) and microions (small red and blue spheres) dispersed in a dielectric continuum 
solvent and (bottom) cell model of a surface-charged microgel of swollen radius $a$ confined
to a spherical cell of radius $R$ with counterions (red) and coions (blue).}\label{figmodel}
\end{figure}

Theoretical descriptions of polyelectrolyte solutions and charge-stabilized colloidal suspensions
are often based on the primitive model, which approximates the solvent as an implicit medium 
-- a dielectric continuum characterized entirely by a dielectric constant $\epsilon$ -- and 
considers explicitly only the charged species~\cite{deserno-holm2001, denton-cecam2014}. 
Within this coarse-grained model, we consider an aqueous suspension of $N_m$ ionic microgels,
each composed of a microscopic cross-linked network of polymer chains, dispersed in a solvent 
(water) of volume $V$ at temperature $T$, as depicted in Fig.~\ref{figmodel}.

We model ionic microgels as spherical macroions of dry radius $a_0$, swollen radius $a$, 
and fixed charge $-Ze$, resulting from dissociation of $Z$ monovalent counterions 
(of charge $e$) from the polymer chains.
The swollen microgels are assumed to be permeable to solvent, counterions, and coions 
(see Fig.~\ref{figmodel}). Assuming that a microgel in its dry (unswollen) state 
comprises randomly close-packed spherical monomers, the number of monomers per microgel 
is related to the monomer radius $r_{\rm mon}$ by $N_{\rm mon}=0.63(a_0/r_{\rm mon})^3$. 

In experiments, the distribution of fixed charge can be controlled, to some extent, 
via the chemical synthesis protocol. 
In some systems, e.g., poly-N-isopropilacrylamide (pNIPAM) microgels,
the fixed charges originate from the initiator species for the polymerization reaction 
in the synthesis, which are believed to end up localized near 
the particle surface~\cite{nieves-pnas2016,scotti-binary-pre2017,gasser2019,Zhou2012}. 
We consider here the ideal case that the fixed charge is strictly confined to and 
uniformly distributed over the particle surface, with number density 
described by a Dirac delta-function,
\begin{equation}
n_f(r) =\frac{Z}{4\pi a^2}\,\delta(r-a),
\label{nf}
\end{equation}
where $Z$ is the microgel valence and $r$ is the radial distance from the microgel center. 
Elsewhere~\cite{Wypysek-Scotti2019,alziyadi-denton2021}, we investigated swelling of ionic 
microcapsules, whose fixed charge is spread throughout a spherical shell of a hollow microgel.
In contrast, the microgels studied here are not hollow, but are completely filled with hydrogel.
The microions are modeled as monovalent point charges of valence $z_{\pm}=\pm 1$. 
In Donnan equilibrium with a 1:1 electrolyte (salt) reservoir, 
the suspension contains $N_s$ dissociated salt ion pairs, which contribute equal numbers 
of coions and additional counterions to the solution. Electroneutrality of the suspension 
dictates the total numbers of counterions and coions as $N_+=Z+N_s$ and $N_-=N_s$, 
respectively, for a total of $N_{\mu}=Z+2N_s$ microions.

\subsection{Cell Model}
The cell model, originally proposed by Wigner and Seitz to calculate electronic properties 
of solids~\cite{ashcroft76,kittel2004}, can also be applied to polyelectrolyte solutions,
as first recognized by Marcus~\cite{marcus1955}, charge-stabilized colloidal 
suspensions~\cite{deserno-holm2001}, and hydrogels~\cite{holm2009}.
In this context, the cell model reduces a system of many macroions dispersed in an electrolyte 
solution to a single macroion confined to a cell of like shape, along with stoichiometric
numbers of counterions and coions (Fig.~\ref{figmodel}). This relatively simple model, by
explicitly including microions, retains microion-microion and microion-macroion interactions,
but sacrifices macroion-macroion interactions and correlations, aside from a mean-field 
contribution to the total electrostatic energy.

In the cell model of an ionic microgel suspension, a swollen, spherical microgel (macroion)
of radius $a$, with radially symmetric fixed charge density $n_{f}(r)$, is centered in a 
spherical cell of radius $R$, which determines the suspension volume fraction $\phi=(a/R)^3$. 
From spherical symmetry, the electrostatic potential and microion densities $n_{\pm}(r)$
depend on only the radial distance $r$ from the center of the cell. Further, electroneutrality 
requires that the electric field vanishes at the cell boundary ($r=R$). 

The model system is governed by a Hamiltonian $H$, which we assume can be 
separated~\cite{weyer-denton2018}, according to
\begin{equation}
H=H_g+H_e,
\label{H}
\end{equation}
into a ``gel" Hamiltonian $H_g$, associated with mutual interactions among monomers of the 
cross-linked polymer chains and the solvent molecules, and an ``electrostatic" Hamiltonian 
$H_e$, which includes the Coulomb energies of interaction between ions (fixed and mobile).
The gel Hamiltonian incorporates elastic and mixing degrees of freedom of a polymer network 
in solution. The electrostatic Hamiltonian further decomposes, according to
\begin{equation}
H_e= U_m(a)+U_{m\mu}(\{\br\};a)+U_{\mu\mu}(\{\br\}),
\label{He}
\end{equation}
whose three terms account, respectively, for the electrostatic self-energy of a microgel 
and macroion-microion and microion-microion pair interaction energies, which depend on 
the microion coordinates, $\{{\br}\}=\{\br_{1}, . . . , \br_{N_{\mu}}\}$. 
Note that the first two terms on the right side of Eq.~(\ref{He}) are functions of 
the microgel radius, and thus vary with swelling of the polymer network.

A surface-charged spherical macroion has electrostatic self-energy, in thermal ($k_BT$) units,
\begin{equation}
\beta U_m(a)=Z^2\frac{\lambda_B}{2a},
\label{Um}
\end{equation}
where the Bjerrum length, $\lambda_B\equiv\beta e^2/\epsilon$, is the distance at which two 
elementary charges $e$ have a Coulomb energy equal to the thermal energy $k_BT\equiv 1/\beta$. 
The microgel-microion interaction energy can be expressed as
\begin{equation}
U_{m\mu}(\{\br\};a)=\sum_{i=1}^{N_+}v_{m+}(\br_{i};a)+\sum_{i=1}^{N_-}v_{m-}(\br_{i};a),
\label{Ummu}
\end{equation}
where $v_{m\pm}$ are the macroion-counterion and macroion-coion electrostatic pair potentials:
\begin{equation}
\beta v_{m\pm}(r;a) = \left\{ \begin{array}{cc} \vspace{2mm} 
\mp Zz\lambda_B/r, & \hspace{2mm} r>a\\ \vspace{2mm} 
\mp Zz\lambda_B/a, & \hspace{2mm} r\leq a.\\              
\end{array} \right.
\label{vpm}
\end{equation}
The fact that the macroion-microion pair potentials depend on the microgel radius $a$ has 
direct implications for swelling of ionic microgels, as discussed below in Sec.~\ref{theory}.
In passing, we note that, in the full primitive model of a suspension of many microgels, 
the electrostatic Hamiltonian would include, in addition to the three terms in Eq.~(\ref{He}), 
a fourth term accounting for microgel-microgel electrostatic interactions,
\begin{equation}
U_{mm}(a)=\sum_{i<j=1}^{N_m}v_{mm}(r_{ij}; a).
\label{vpm}
\end{equation}
For spherical microgels, however, the dependence of the bare Coulomb microgel-microgel 
pair potential $v_{mm}(r;a)$ on microgel radius arises only when the microgels overlap.
For other shapes of microgels, pair interactions do not have this symmetry property,
and would depend on the orientations of the particles at all separations. 

In the cell model, microgel-microgel electrostatic interactions are
implicitly accounted for through the boundary condition that the electric field vanishes 
at the cell edge. The focus on a single microgel of course necessitates neglect of 
microgel-microgel correlations. In an alternative approach, microion degrees of freedom 
are averaged over (traced out of the partition function), reducing the multi-component 
mixture to an effective one-component model of pseudo-macroions that interact via an 
{\it effective} screened-Coulomb (Yukawa) 
pair potential~\cite{denton2003,gottwald2005,weyer-denton2018}. In this multi-centered
model, effective pair interactions influence swelling at densities well below close packing, 
corresponding to nearest-neighbor separations comparable to the Debye screening length. 
Remarkably, as shown in ref.~\cite{weyer-denton2018} for volume-charged microgels, the cell 
and one-component models predict very similar swelling behavior up to close-packing densities.

\section{Theory of Microgel Swelling}\label{theory}
Swelling of microgels is governed by the single-microgel osmotic pressure $\pi_m$, 
defined as the difference in pressure between the interior and exterior of the microgel.
This single-microgel property is to be distinguished from the osmotic pressure of
a microgel suspension $\pi_s$, which is a bulk (macroscopic) property.
In a canonical ensemble description, in which the suspension is assumed to be closed (i.e.,
has a fixed number of particles), the single-microgel osmotic pressure is thermodynamically 
defined as a derivative of the Helmholtz free energy per microgel $F$ with respect to 
the microgel volume $v_m=4\pi a^3/3$. Within the spherical cell model, 
\begin{equation}
\pi_m=-\left(\frac{\partial F}{\partial v_m}\right)_{N_{\mu}, R, T}
=-\frac{1}{4\pi a^2}\left(\frac{\partial F}{\partial a}\right)_{N_{\mu}, R, T},
\label{pim2}
\end{equation} 
revealing the connection between $\pi_m$ and the variation of the free energy with 
swollen radius. In contrast, the suspension osmotic pressure depends on the variation 
of the free energy with cell volume or radius, 
\begin{equation}
\pi_s=-\left( \frac{\partial F}{\partial V} \right)_{N_{\mu}, T}
=-\frac{1}{4\pi R^2}\left( \frac{\partial F}{\partial R} \right)_{N_{\mu}, T}.
\label{pis}
\end{equation} 
The cell theorem~\cite{marcus1955,Wennerstrom} further relates $\pi_s$ to the 
total microion density at the cell boundary:
\begin{equation}
\beta\pi_s=n_+(R)+n_-(R).
\label{cell-theorem}
\end{equation} 
In passing, we note that the results derived below can be equally well obtained within 
the semi-grand canonical ensemble, in which the suspension is free to exchange microions 
with a salt reservoir, which describes the condition of Donnan 
equilibrium~\cite{denton-tang2016,denton-alziyadi2019}.

Separation of the Hamiltonian into electrostatic and gel terms [Eq.~(\ref{H})] implies 
that the canonical partition function ${\cal Z}$ factorizes, according to 
${\cal Z}={\cal Z}_e{\cal Z}_g$, into electrostatic and gel factors. 
Within the spherical cell model, the electrostatic partition function takes the form
\begin{equation}
{\cal Z}_e(N_{\mu}, a, R, T)\propto \prod_{i=1}^{N_{\mu}}\int_0^R dr_i\,r_i^2 e^{-\beta H_e},
\label{Ze}
\end{equation} 
where the position integrals cover possible configurations of all microions within the cell. 
Factorization of the partition function implies decomposition of the free energy,
\begin{equation}
F=-k_BT\ln{\cal Z}=F_g+F_e,
\label{F}
\end{equation}
into (1) a gel contribution, $F_g=-k_BT\ln{\cal Z}_g$, associated with short-range 
monomer-monomer interactions, mixing entropy of the polymer-solvent mixture, and 
conformational entropy of the polymer chains, and (2) an electrostatic contribution,
$F_e=-k_BT\ln{\cal Z}_e$, due to long-range (Coulomb) ion-ion interactions. 
Correspondingly, the single-microgel osmotic pressure separates, via
\begin{equation}
\pi_m(\alpha)=\pi_g(\alpha)+\pi_e(\alpha),
\label{pim1}
\end{equation} 
into gel and electrostatic contributions, $\pi_g(\alpha)$ and $\pi_e(\alpha)$, both 
being functions of the microgel linear swelling ratio $\alpha\equiv a/a_0$.

In thermodynamic equilibrium, a microgel swells until its osmotic pressure vanishes,
i.e., $\pi_m(\alpha)=0$. As we showed in previous work~\cite{denton-alziyadi2019},
this condition ensures continuity of the radial component of the total pressure tensor 
across the microgel surface in equilibrium. Note that the individual contributions 
to $\pi_m$ may be nonzero, as long as their sum vanishes in equilibrium. 

As an approximation for the gel contribution to the free energy, we adopt the mean-field 
Flory-Rehner theory of polymer networks~\cite{P.J.Flory,flory-rehner1943-I,flory-rehner1943-II},
which combines polymer-solvent interactions and mixing entropy with elastic free energy of the 
network. For a microgel of swollen radius $a$ and dry radius $a_0$, composed of $N_{\rm mon}$ 
monomers and $N_{\rm ch}$ distinct chains (cross-linked into a network) in a solvent, 
the gel free energy takes the form
\begin{eqnarray}
\beta F_g(\alpha)&=&N_{\rm mon}[(\alpha^3-1)\ln(1-\alpha^{-3})+\chi(1-\alpha^{-3})]
\nonumber\\[1ex]
&+&\frac{3}{2}N_{\rm ch}(\alpha^2-\ln\alpha-1),
\label{Fg}
\end{eqnarray}
where $\chi$ is the Flory solvency parameter, associated with polymer-polymer, 
polymer-solvent, and solvent-solvent interactions.
The corresponding gel contribution to the single-microgel osmotic pressure can be 
written as 
\begin{eqnarray}
\beta\pi_g(\alpha)v_m&=&-N_{\rm mon}[\alpha^3\ln(1-\alpha^{-3})+\chi\alpha^{-3}+1]
\nonumber\\[1ex]
&-&N_{\rm ch}(\alpha^2-1/2).
\label{pig}
\end{eqnarray}

The electrostatic contribution to the single-microgel osmotic pressure can be 
derived in principle from the electrostatic contribution to the free energy 
via~\cite{colla-likos2014}
\begin{equation}
\pi_e=-\frac{1}{4\pi a^2}\left(\frac{\partial F_e}{\partial a}\right)_{N_{\mu}, R, T}.
\label{pie0}
\end{equation} 
However, this expression is only as accurate as the approximation for the free energy.
Alternatively, from Eqs.~(\ref{Um}) and (\ref{Ummu}), $\pi_e$ can be expressed as
\begin{equation}
\pi_e=-\frac{1}{4\pi a^2}\left( \frac{\partial U_m(a)}{\partial a}
+\la\frac{\partial U_{m\mu}(a)}{\partial a}\ra\right)_{N_{\mu},R,T},
\label{pie1}
\end{equation}
where angular brackets denote an ensemble average over configurations of the microions.
Applying the latter approach to ionic microgels requires specifying a model for the fixed 
charge distribution. Assuming a surface-charged microgel and using the appropriate 
macroion-microion interaction potentials [Eq.~(\ref{vpm})], the second term on the right 
side of Eq.~(\ref{pie1}) takes the explicit form
\begin{equation}
\beta\la\frac{\partial U_{m\mu}(a)}{\partial a}\ra=\frac{Z\lambda_B}{a^2}
\la\sum_{i(r_i\leq a)}z_i\ra,
\label{dUmmuda}
\end{equation}
where the sum over $i$ ($i=\pm$ for counterions or coions) includes only microions
inside the microgel. Finally, substituting the self energy [Eq.~(\ref{Um})] 
and the expression from Eq.~(\ref{dUmmuda}) into Eq.~(\ref{pie1}), we obtain the
electrostatic component of the single-microgel osmotic pressure:
\begin{equation}
\beta\pi_e v_m=\frac{Z\lambda_B}{3a}\left(\frac{Z}{2}
-\la{N_+}\ra+\la{N_-}\ra\right),
\label{pie2}
\end{equation}
where 
\begin{equation}
\la N_{\pm}\ra =4\pi\int_{0}^{a}dr\, r^2 n_{\pm}(r)
\label{Npm}
\end{equation}
represent mean numbers of counterions and coions {\it inside} the microgel, 
ensemble averaged over microion configurations~\cite{denton-alziyadi2019}. 
The first term on the right side of Eq.~(\ref{pie2}) derives from the microgel
self-energy, while the remaining terms are associated with the microion distribution. 
Equations~(\ref{pim1}), (\ref{pig}), and (\ref{pie2}), combined with the 
condition $\pi_m(\alpha)=0$, determine a microgel's equilibrium swollen size.
In previous work~\cite{denton-alziyadi2019}, we showed that our expression for
the electrostatic component of the single-microgel osmotic pressure in the cell model 
is equivalent to the jump at the microgel surface in the radial component 
of the electrostatic pressure tensor~\cite{trizac-hansen1997,Widom2009}.

Note that Eq.~(\ref{pie2}), akin to the cell theorem for the suspension osmotic 
pressure [Eq.~(\ref{cell-theorem})], is an {\it exact} theorem within the cell model,
fully accounting for microion correlations. In multi-center models of microgel 
suspensions, these theorems cease to be exact, as they neglect osmotic pressure 
contributions from microgel-microgel pair interactions and correlations.
Nevertheless, since these pair interactions depend on microgel size only for overlapping 
microgels, Eq.~(\ref{pie2}) gives a reasonable approximation for dilute suspensions,
and should be accurate even for concentrations approaching 
close-packing~\cite{weyer-denton2018}.

\section{Computational Methods}\label{methods}
\subsection{Poisson-Boltzmann Theory in the Cell Model}
Practical implementation of the theory described in Sec.~\ref{theory} to model equilibrium 
swelling of ionic microgels requires estimates for the mean numbers of counterions and 
coions inside a microgel. Within the cell model, a convenient method for computing 
$\la N_{\pm}\ra$ is Poisson-Boltzmann (PB) theory. Based on a mean-field approximation that 
neglects microion-microion correlations, this theory can be rigorously derived, e.g., 
via calculus of variations~\cite{denton-alziyadi2019} and
classical density-functional theory~\cite{lowen92}.

Defining $\psi(r)\equiv\beta e\phi(r)$ as the reduced (dimensionless) form of the
electrostatic potential $\phi(r)$, where $r$ is the radial distance from the center 
of the cell, the Poisson equation can be expressed as
\begin{equation}
\nabla^2\psi(r)=-4\pi\lambda_B[n_+(r)-n_-(r)-n_f(r)],
\label{Poisson}
\end{equation} 
where the right-hand side includes the total charge density, including the mobile microions 
and the fixed charge on the microgel. For a suspension in Donnan equilibrium 
with a salt reservoir of average salt density $n_0$, the mean-field Boltzmann approximation 
for the microion equilibrium densities takes the form
\begin{equation}
n_{\pm}(r)=n_0\,\text{exp}[\mp\psi(r)].
\label{Boltzmann}
\end{equation} 
Combining Eqs.~(\ref{Poisson}) and (\ref{Boltzmann}) yields the (nonlinear) 
Poisson-Boltzmann equation:
\begin{equation}
\psi''(r)+\frac{2}{r}\psi'(r)=\kappa^2\sinh\psi(r)+4\pi\lambda_B n_f(r),
\label{pb1}
\end{equation} 
where $\kappa\equiv\sqrt{8\pi\lambda_Bn_0}$ is the Debye screening constant. 
For a closed, salt-free suspension, with fixed counterion density,
the PB equation takes the alternate form
\begin{equation}
\psi''(r)+\frac{2}{r}\psi'(r)=-\tilde\kappa^2\exp[-\psi(r)]+4\pi\lambda_B n_f(r),
\label{pb2}
\end{equation} 
where $\tilde\kappa\equiv\sqrt{4\pi\lambda_B\tilde n_+}$ is the effective screening constant 
within the suspension and $\tilde n_+$ is the counterion number density at $\psi=0$.

In the case of a microgel with fixed charge confined to its surface, 
we substitute Eq.~(\ref{nf}) into Eq.~(\ref{pb1}) to obtain
\begin{equation}
\psi''(r)+\frac{2}{r}\psi'(r)=\kappa^2\sinh\psi(r)+\frac{Z\lambda_B}{a^2}\, \delta(r-a),
\label{pb3}
\end{equation}
which is the PB equation in the spherical cell model for a suspension of 
surface-charged microgels in Donnan equilibrium.
This equation is subject to the boundary conditions that the electric field must vanish 
at the center of the cell, $\psi'(0)=0$ (by symmetry), and on the cell boundary, 
$\psi'(R)=0$ (by electroneutrality). Applying these boundary conditions, we 
numerically solved Eq.~(\ref{pb3}) for $\psi(r)$ inside and outside the microgel 
and matched the solutions in the two regions at the microgel surface to ensure 
continuity of the potential and the correct discontinuity of the electric field at $r=a$:
\begin{equation}
\lim_{\delta\to 0}\left[\psi'(a+\delta)-\psi'(a-\delta)\right]=\frac{Z\lambda_B}{a^2}.
\label{psi-prime}
\end{equation}
From our solution for $\psi(r)$, we computed the microion density profiles 
from Eq.~(\ref{Boltzmann}) and the electrostatic component of the single-microgel 
osmotic pressure from Eqs.~(\ref{pie2}) and (\ref{Npm}).

Previously, we showed~\cite{denton-alziyadi2019} that a careful analysis of the 
pressure tensor within the cell model implementation of PB theory relates the 
electrostatic component of the single-microgel osmotic pressure to the discontinuity
in the radial component of the electrostatic pressure tensor at the microgel surface:
\begin{equation}
\beta\pi_e=\frac{Z}{8\pi a^2}\lim_{\delta\to 0}\left|\psi'(a+\delta)+\psi'(a-\delta)\right|.
\label{delta-P-mechanical-microgel-surface-charge}
\end{equation}
We further showed that this expression is equivalent to our exact statistical mechanical 
relation [Eq.~(\ref{pie2})].

\subsection{Molecular Dynamics Simulations}
An independent means of computing the microion density distributions is molecular simulation. 
To validate our PB theory predictions, we implemented and performed canonical ensemble
molecular dynamics (MD) simulations to compute $\la N_{\pm}\ra$ within the cell model. 
Using the LAMMPS package~\cite{lammps,Plimpton1995}, we simulated a collection of counterions 
(and coions), interacting via a hybrid pair potential, combining a long-range Coulomb pair 
potential with a short-range, repulsive, cut-off Lennard-Jones pair potential. To mimic the 
influence of a central, spherical ionic microgel, we applied a radial 
``external" force~\cite{denton-tang2016},
\begin{equation}
F_{\rm ext}(r)=\mp\frac{Ze^2}{4\pi\epsilon}
\left\{ \begin{array}
{l@{\quad}l}
r/a^3,
& r\le a, \\[2ex]
1/r^2,
& r>a, \end{array} \right.
\label{Fext}
\end{equation}
which is attractive/repulsive ($\mp$) for counterions/coions. 

The microions were initialized on the sites of an FCC lattice with random
velocities consistent with a constant temperature, and confined to a spherical region 
by a repulsive Lennard-Jones wall force. During the simulations, trajectories of the 
mobile microions were computed by numerically integrating Newton's equations of motion 
using the velocity-Verlet algorithm. The average temperature was held constant via a 
Nos\'e-Hoover thermostat. After equilibrating the system for $10^6$ time steps (fs),
we continued each run for another $10^8$ steps, during which time we collected statistics.

\begin{figure}[h]
\includegraphics[width=0.9\columnwidth,angle=0]{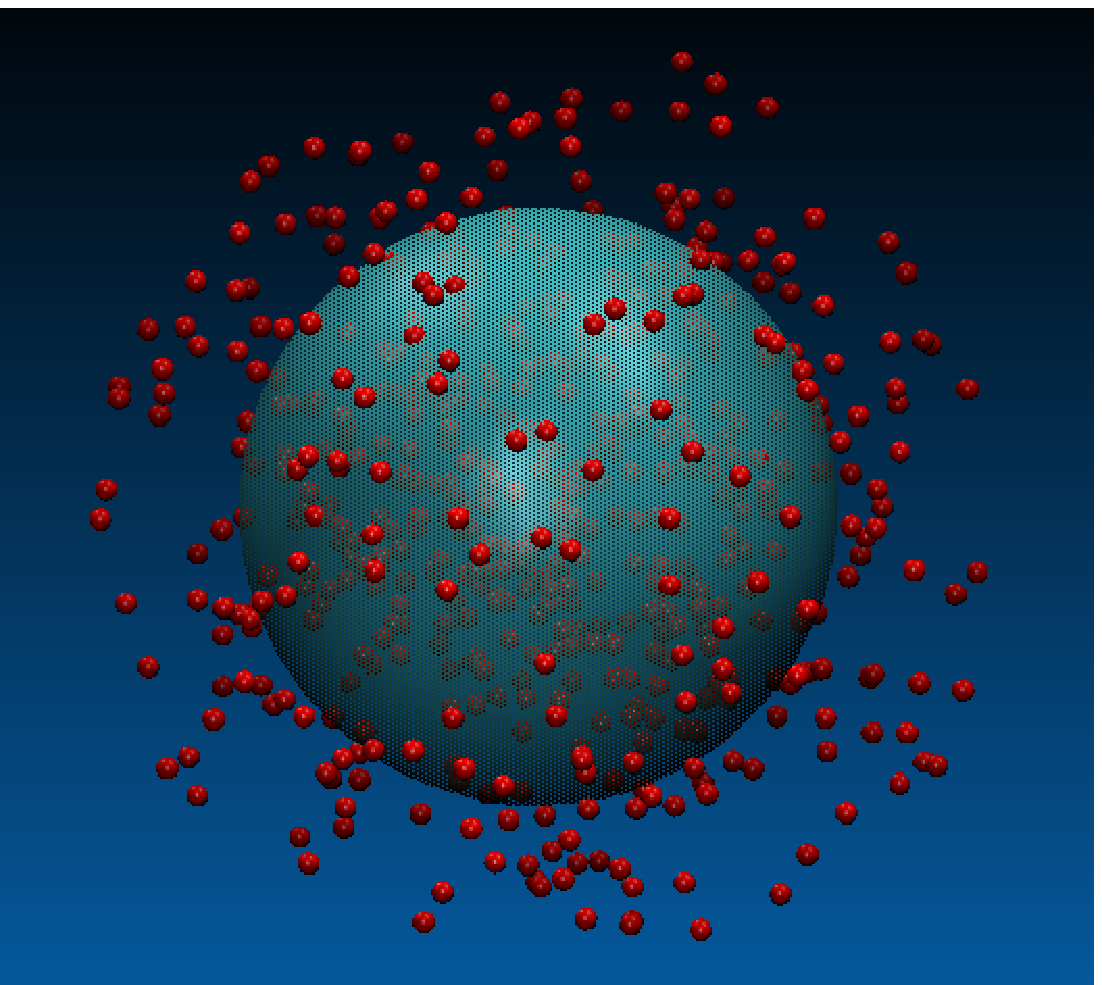}
\vspace*{-0.2cm}
\caption{Snapshot from MD simulation of an ionic microgel (blue sphere) and mobile counterions
(red spheres).}
\label{figsn}
\end{figure}

In all MD simulations and PB calculations, the following system parameters were kept fixed:
Bjerrum length $\lambda_B$, cell radius $R$, microgel dry radius $a_0$, cross-link fraction, 
dry volume fraction $\phi_0$, valence $Z$, and Flory solvency parameter $\chi$.
For a given value of $Z$, the total number of dissociated monovalent counterions 
was $N_+=Z$ to maintain electroneutrality. Figure~\ref{figsn} shows a typical 
snapshot from one of our simulations of an ionic microgel and counterions.

\section{Results and Discussion}\label{results}
To explore the dependence of equilibrium swelling behavior of ionic microgels on fixed 
charge distribution, we implemented the models, theory, and computational methods 
described in Secs.~\ref{models}-\ref{methods}. We consider here aqueous suspensions 
at temperature $T=293$ K, with a uniform Bjerrum length, $\lambda_B=0.714$ nm, 
assuming equal dielectric constants inside and outside of the microgels. As representative 
system parameters, we chose a dry microgel radius of $a_0=10$ nm, corresponding to 
$N_{\rm mon}=2\times 10^5$ monomers of radius $r_{\rm mon}\simeq 0.15$ nm; 
number of chains $N_{\rm ch}=100$, corresponding to average cross-link fraction 
$x=0.5N_{\rm ch}/N_{\rm mon}=2.5\times 10^{-4}$; valence $Z=1000$; and 
Flory solvency parameter $\chi=0.5$.

In previous work~\cite{denton-tang2016}, we developed a similar modeling approach and applied 
it to spherical ionic microgels carrying fixed charge uniformly distributed over their volume. 
We numerically solved the PB equation and performed MD simulations to compute average
microion number density profiles, and thus average numbers of interior microions 
$\la N_{\pm}\ra$. From the resulting electrostatic component of the single-microgel osmotic
pressure $\pi_e$, we determined equilibrium swelling ratios. Here, we apply a comparable
approach to microgels with fixed charge uniformly distributed over their surfaces. 

\begin{figure}[h]
\includegraphics[width=0.9\columnwidth]{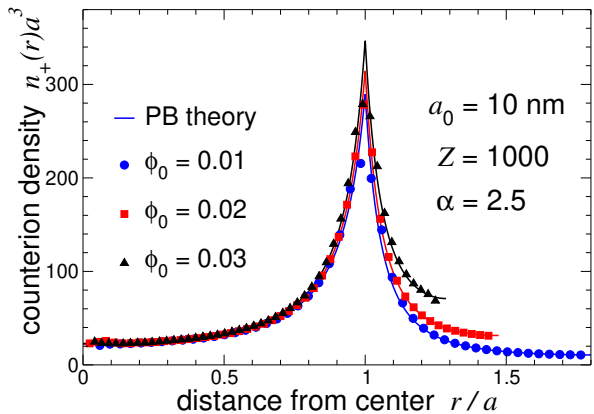}
\vspace*{-0.2cm}
\caption{Counterion number density profiles around a surface-charged ionic microgel 
of dry radius $a_0=10$ nm, swollen radius $a=25$ nm, and valence $Z=1000$ in a salt-free
aqueous solvent at room temperature ($\lambda_B=0.714$) in the spherical cell model. 
For dry volume fractions $\phi_0=0.01$, 0.02, and 0.03, predictions of PB theory 
(solid curves) agree closely with MD simulation data (symbols). The corresponding
swollen volume fractions are $\phi=0.15625$, 0.3125, and 0.46875.
}
\label{profile}
\end{figure}

Figure~\ref{profile} shows our results for average counterion number density profiles 
inside and outside of a surface-charged microgel of swollen radius $a=25$ nm 
(swelling ratio $\alpha=2.5$) in the spherical cell model with no salt. Excellent 
agreement between predictions of PB theory and data from our MD simulations over a 
range of dry volume fractions serves to validate our implementations of both methods. 
Consistent with the boundary conditions on the PB equation for the electrostatic potential,
the counterion density profiles are relatively flat near the microgel center ($r=0$) and the 
cell edge ($r=R$), where the electric field vanishes. Furthermore, the counterion density 
peaks at the surface of the microgel, where the electric field is highest. The cusp in 
the $n_+(r)$ curve at $r=a$ is a consequence of the strict confinement of the fixed charge 
to the particle surface. 
 
With increasing dry volume fraction, the maxima of the $n_+(r)$ curves (at $r=a$) increase.
This trend is to be expected, since the counterion density closely tracks the electrostatic
potential [Eq.~(\ref{Boltzmann})], whose magnitude $|\psi(r)|$ attains a maximum at the 
surface and whose surface value $|\psi(a)|$ increases with increasing $\phi_0$. This increase
in peak height reflects a stronger attraction of counterions to the surface at higher
microgel volume fractions. Furthermore, the average number of interior counterions $\la N_+\ra$
[Eq.~(\ref{Npm})] also increases with increasing $\phi_0$, a trend that results largely 
from the corresponding decrease in exterior volume. Interestingly, inside and away from the 
microgel surface, the counterion density profile changes relatively little.
This behavior is expected, as the environment of the interior counterions is weakly
affected by a change in volume outside of the microgel as $\phi_0$ changes.
Note that at low volume fractions (dilute conditions) the counterion density near the
microgel center is higher than at the cell edge, while at higher volume fractions
this relation is reversed.
In previous work~\cite{Wypysek-Scotti2019,alziyadi-denton2021}, we found that 
distributing the fixed charge over the shell thickness of a spherical microcapsule 
smooths and broadens the counterion density profile. 

\begin{figure}[h]
\includegraphics[width=0.9\columnwidth]{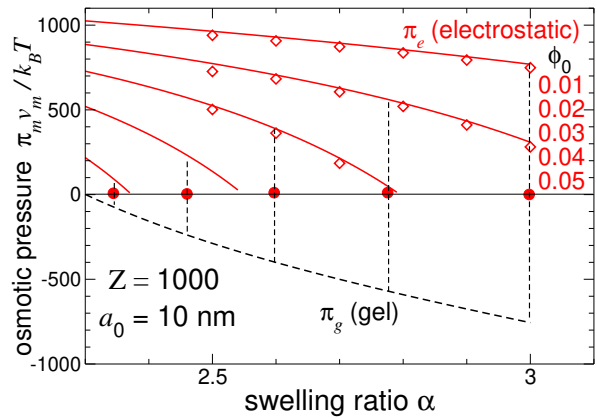}
\vspace*{-0.2cm}
\caption{Electrostatic and gel components of single-microgel osmotic pressure 
vs.~swelling ratio $\alpha$ for surface-charged microgels in salt-free aqueous
suspensions of dry volume fraction $\phi_0=0.01-0.05$ in the spherical cell model 
for same system parameters as in Fig.~\ref{profile}. The electrostatic component 
$\pi_e$ from PB theory (solid red curves) agrees closely with MD simulation data 
(open red symbols). The gel component $\pi_g$ (dashed black curve) is computed from 
Flory-Rehner theory [Eq.~(\ref{pig})]. At equilibrium swelling, the total microgel 
osmotic pressure vanishes: $\pi_m=\pi_e+\pi_g=0$ (filled symbols).
}
\label{plot-pe}
\end{figure}

As discussed in Sec.~\ref{methods}, the electrostatic component of the 
single-microgel osmotic pressure $\pi_e$ involves contributions both from the self-energy 
of the microgel fixed charge and from the density distributions of the mobile microions.
Numerically integrating the microion density profiles over the volume of the swollen 
microgel yields the average numbers of interior microions $\la N_{\pm}\ra$ [Eq.~(\ref{Npm})]. 
In the salt-free case, substituting the average number of interior counterions 
$\la N_+\ra$ into Eq.~(\ref{pie2}) gives $\pi_e$ for a microgel of a given swollen radius.
Repeating these calculations for different swollen radii, we computed $\pi_e$
over a range of microgel swelling ratios and dry volume fractions. Figure~\ref{plot-pe} 
shows our results for $\pi_e$ vs.~$\alpha$ for several values of $\phi_0$, together 
with the gel component of the single-microgel osmotic pressure [Eq.~(\ref{pig}], which is 
independent of $\phi_0$. Again, our predictions from PB theory are in very close agreement 
with our simulation data, as is to be expected, since $\pi_e$ is directly determined by 
the value of $\la N_+\ra$. 

For a given microgel swollen radius, as the dry volume fraction increases (from $\phi_0=0.01$ 
to 0.05), the electrostatic component of the single-microgel osmotic pressure steadily 
decreases (red curves and symbols in Fig.~\ref{plot-pe}). This trend can be attributed to 
the progressive neutralization of ionic microgels with increasing concentration of the 
suspension, as counterions become increasingly confined within the microgel,
i.e., as $\la N_+\ra$ increases.
(Note that in the cell model, increasing $\phi_0$ implies decreasing cell radius $R$.) 
Furthermore, for these system parameters, at a given concentration, $\pi_e$ also 
monotonically decreases with increasing swelling ratio (i.e., with increasing
swollen radius for a fixed cell radius). 
At sufficiently large $\alpha$, the self-energy and counterion contributions 
to the electrostatic osmotic pressure balance and $\pi_e=0$. Thus, concentrating the
suspension (increasing $\phi_0$) or enlarging the microgels (increasing $\alpha$) weakens
electrostatic effects (lowering $\pi_e$), which tends to promote deswelling and expulsion
of solvent. Conversely, decreasing $\phi_0$ or $\alpha$ enhances electrostatic effects, 
which favors swelling and absorption of solvent.

The gel component of the single-microgel osmotic pressure $\pi_g$ [Eq.~(\ref{pig})], 
associated with mixing entropy and elasticity, counters the electrostatic component.
Recall that the total single-microgel osmotic pressure $\pi_m$ -- the sum of the 
electrostatic and gel components [Eq.~(\ref{pim1})] -- vanishes when a microgel attains 
its equilibrium swollen size. An uncharged (nonionic) microgel is in equilibrium when
$\pi_g$ (dashed black curve in Fig.~\ref{plot-pe}) itself vanishes (at $\alpha\simeq 2.3$
in this case). With increasing swelling ratio, $\pi_g$ monotonically decreases, becoming 
negative when $\alpha>2.3$, but with a qualitatively different $\alpha$ dependence. 
While a nonionic microgel would be out of equilibrium beyond this swelling ratio, 
an ionic microgel can be stabilized by the electrostatic component of the osmotic pressure, 
here arising from the fixed surface charge.
The red dots in Fig.~\ref{plot-pe} mark the equilibrium size of the microgel,
at which $\pi_g(\alpha)+\pi_e(\alpha)=0$, for different dry radii.

\begin{figure}[h]
\includegraphics[width=0.9\columnwidth]{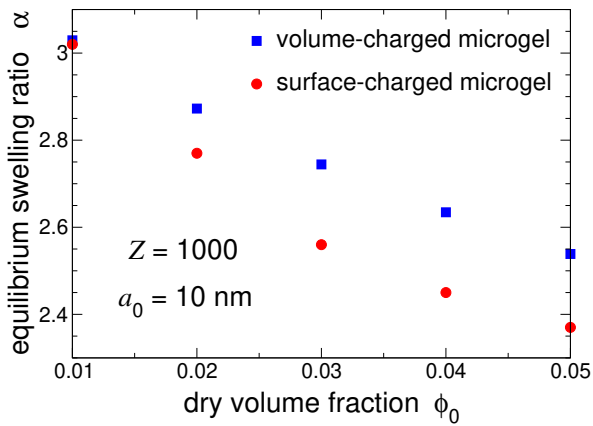}
\vspace*{-0.2cm}
\caption{Equilibrium swelling ratio $\alpha$ vs.~dry volume fraction $\phi_0$
for salt-free aqueous suspensions of surface-charged (red circles) and 
volume-charged (blue squares) microgels in the spherical cell model for same 
system parameters as in Fig.~\ref{profile}, 
illustrating stronger deswelling of surface-charged microgels.
Over this range of $\phi_0$, the swollen volume fraction, $\phi=\alpha^3\phi_0$,
varies (nonlinearly) from about 0.28 to 0.67 for surface-charged microgels.}
\label{surface-volume}
\end{figure}

The resulting equilibrium swelling behavior is summarized in Fig.~\ref{surface-volume}, 
which shows progressive deswelling with increasing $\phi_0$. 
Over this range of dry volume fraction, the swollen volume fraction, $\phi=\alpha^3\phi_0$,
varies (nonlinearly) from about 0.28 to 0.67 for surface-charged microgels, 
remaining below close-packing of spheres.
Thus, the deswelling of these ionic microgels is driven 
mainly by electrostatic effects, rather than by steric interparticle interactions.
With increasing $\phi_0$, increasingly more counterions reside, on average, 
inside the microgel, lowering the contribution of microgel-counterion electrostatic 
interactions to the osmotic pressure [Eq.~(\ref{dUmmuda})], and thus inducing deswelling.

Our predictions of progressive deswelling of ionic microgels with increasing 
concentration are qualitatively consistent with experimental observations of deswelling, e.g.,
of pNIPAM and poly(vinylpyridine) microgels~\cite{nieves-prl2015,nieves-pnas2016,
scotti-binary-pre2017,gasser2019,Zhou2012,scotti2021}. In particular, pNIPAM microgels, 
whose fixed charge is localized near the particle periphery, similar to our surface-charged 
model, exhibit such a deswelling response~\cite{nieves-pnas2016,scotti-binary-pre2017,
gasser2019,Zhou2012}. A direct quantitative comparison between theory and experiment may be
possible, but is complicated by the challenge of precisely measuring the fixed charge 
and by the neglect in our model of any variation of fixed charge with microgel concentration.

Fern{\'a}ndez-Nieves {\it et al.}~\cite{nieves-pnas2016,scotti-binary-pre2017,gasser2019}
have suggested a model of ionic microgels in which counterions that are closely associated 
with fixed charge, i.e., quasi-bound to the surface with an electrostatic potential energy 
$\gtrsim k_BT$, may, at sufficiently high microgel concentration, be released (i.e., become free) 
and thereby contribute to the osmotic pressure of the suspension. Such a release of counterions 
would alter the effective fixed charge of the microgels. Since our coarse-grained, cell-model-based 
approach focuses, for simplicity, on a single microgel carrying a fixed charge that is independent of 
system properties, it does not directly describe release of quasi-bound counterions in response 
to changes of concentration. A multi-center model of microgel suspensions that includes
microgel-microgel interactions may be a basis for incorporating a concentration-dependent 
fixed charge, using concepts of charge renormalization with a thermal definition of effective 
fixed charge~\cite{denton2010}. While such an approach is beyond the scope of the present study,
it may be worth exploring in future. Nevertheless, our theoretical predictions and simulations 
do indicate stronger aggregation of counterions near the charged microgel surface with increasing 
concentration (Fig.~\ref{profile}), which might be related to quasi-binding of counterions in 
a more molecular-scale model. 
Figure~\ref{profile} further shows that the counterion number density at the cell edge 
increases with concentration, suggesting increasing overlap of the counterion clouds 
surrounding neighboring microgels, consistent with the conceptual picture presented in
refs.~\cite{nieves-pnas2016,scotti-binary-pre2017,gasser2019}.


We note that our results for counterion number density profiles are qualitatively similar 
to those reported in a related study of surface-charged ionic microgels~\cite{gasser2019},
in which the PB equation was numerically solved in the cell model. However, our conclusions 
regarding equilibrium swelling differ significantly from those in ref.~\cite{gasser2019}, 
where the electrostatic component of the single-microgel osmotic pressure was determined from
the microion number density at the center of the cell,
implying [from Eq.~(\ref{Boltzmann})]
\begin{equation}
\beta\pi_e = n_+(0) + n_-(0) = 2n_0\cosh[\psi(0)].
\label{peGasser}
\end{equation}
The latter approximation, although valid in planar geometry (e.g., for a flat film), 
neglects nonuniformity of the elements of the pressure tensor in curvilinear geometry.
In spherical geometry, for example, the normal (radial) element of the pressure tensor, 
$P_m(r)$, varies with radial distance.  
As described in detail elsewhere~\cite{denton-alziyadi2019,trizac-hansen1997,Widom2009},
this property follows from the requirement that the divergence of the pressure tensor 
vanish in mechanical equilibrium. In particular, for a surface-charged microgel 
in the spherical cell model, the normal element of the electrostatic pressure tensor,
$P_e(r)$, is higher at the microgel surface than at the 
center (see Fig.~14 of ref.~\cite{denton-alziyadi2019}). 
As a consequence, Eq.~(\ref{peGasser}) actually represents the normal element of the
electrostatic pressure at the microgel center, $P_e(0)$, rather than the electrostatic
component of the single-microgel osmotic pressure $\pi_e$. Furthermore, $P_e(0)$ is 
generally significantly lower than $\pi_e$. 

In equilibrium, the normal element of the gel contribution to the pressure tensor, $P_g(r)$, 
must counteract the electrostatic contribution, $P_e(r)$, to mechanically stabilize the 
fixed charge against the electric field ($-\psi'$). For a fixed charge density $n_f(r)$, 
a simple mechanical argument yields the gel contribution as~\cite{denton-alziyadi2019}
\begin{equation}
P_g(r)=-\int_r^R du\, n_f(u)\psi'(u).
\label{Pgr1}
\end{equation}
In general, $P_g(r)=0$ outside the microgel, where $n_f=0$, and is spatially varying 
inside. For our particular model of a surface-charged microgel, with $n_f(r)$ given 
explicitly by a Dirac delta-function [Eq.~(\ref{nf})], Eq.~(\ref{Pgr1}) yields~\cite{denton-alziyadi2019}
\begin{equation}
P_g(r)=-\frac{Z}{8\pi a^2}\lim_{\delta\to 0}[\psi'(a+\delta)+\psi'(a-\delta)]=-\pi_e.
\label{Pgr2}
\end{equation}
Thus, for a surface-charged microgel, $P_g(r)$ jumps discontinuously at the surface 
from zero to a value equal to the negative of the electrostatic component of the 
single-microgel osmotic pressure, and remains constant at that value inside the microgel. 
In contrast, $P_e(r)$ and $P_m(r)$ vary with radial distance due to the spherical geometry.

\begin{figure}[h]
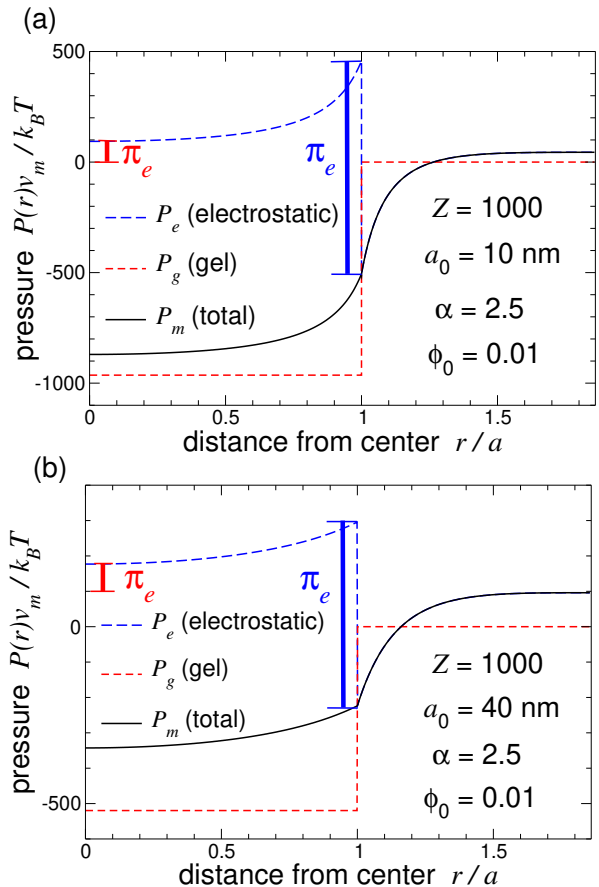

\includegraphics[width=0.9\columnwidth]{per_new_panela.eps}
\includegraphics[width=0.9\columnwidth]{per_new_panelb.eps}
\vspace*{-0.2cm}
\caption{Normal (radial) element of pressure tensor, $P_m(r)$, and electrostatic 
and gel contributions, $P_e(r)$ and $P_g(r)$, respectively, where $P_m(r)=P_e(r)+P_g(r)$, 
for a surface-charged microgel in the spherical cell model with dry volume fraction 
$\phi_0=0.01$, swelling ratio $\alpha=2.5$, and dry and swollen radii (a) $a_0=10$ nm,
$a=25$ nm and (b) $a_0=40$ nm, $a=100$ nm.
Thick red and blue vertical lines have lengths representing the value of $\pi_e$ 
predicted, respectively, by Eq.~(\ref{peGasser})~\cite{gasser2019} and our 
Eq.~(\ref{pie2}) or, equivalently, (\ref{delta-P-mechanical-microgel-surface-charge}).
}
\label{per}
\end{figure}

To illustrate the difference between our approach and that of ref.~\cite{gasser2019},
Fig.~\ref{per} plots the normal element of the pressure tensor in the spherical cell model for 
the case $\phi_0=0.01$, $\alpha=2.5$, separately showing the electrostatic and gel contributions, 
computed using the methods from ref.~\cite{denton-alziyadi2019}. 
Panel (a) is for the case of microgels with dry and swollen radii 
$a_0=10$ nm and $a=25$ nm, respectively; panel (b) is for $a_0=40$ nm and $a=100$ nm, 
which is closer to the experimental system studied in ref.~\cite{gasser2019}. 
The thick blue vertical line has a length equaling our result for $\pi_e$, while the 
thick red vertical line length represents the value predicted by Eq.~(\ref{peGasser}).
Evidently, the two approaches make substantially different predictions.
In fact, when we use Eq.~(\ref{peGasser}), instead of our Eq.~(\ref{pie2}) [equivalently, 
Eq.~(\ref{delta-P-mechanical-microgel-surface-charge})] to compute equilibrium swelling ratios
for the case of $a_0=10$ nm,
we find that $\alpha$ varies only slightly with $\phi_0$, actually increasing 
from 2.355 at $\phi_0=0.01$ to 2.358 at $\phi_0=0.05$. In contrast, our approach, which is 
consistent with nonuniformity of the pressure tensor~\cite{denton-alziyadi2019}, predicts
$\alpha$ decreases from 3.03 at $\phi=0.01$ to 2.37 at $\phi_0=0.05$ for this system
(Fig.~\ref{surface-volume}).

Of particular interest here is a direct comparison between equilibrium swelling of 
surface- and volume-charged microgels. As illustrated in Fig.~\ref{surface-volume}, 
with increasing concentration, surface-charged microgels deswell significantly more
rapidly than volume-charged microgels that have otherwise the same particle properties.
This trend can be attributed in part to the lower self-energy of surface-charged microgels, 
$\beta U_m=Z^2\lambda_B/(2a)$, compared with $\beta U_m=3Z^2\lambda_B/(5a)$ for 
volume-charged microgels~\cite{denton-tang2016}. However, the counterion density distribution 
also clearly plays a role in lowering the electrostatic osmotic pressure, 
which suppresses swelling.

Intermediate between the extremes of surface- and volume-charged microgels
are microgels with the same fixed charge uniformly distributed over the volume of a 
peripheral spherical shell. In previous work on ionic microcapsules 
(hollow microgels)~\cite{Wypysek-Scotti2019,alziyadi-denton2021}, we derived an expression 
for the self-energy of a charged spherical shell with inner and outer radii $R_i$ and $R_o$,
respectively:
\begin{equation}
\beta U_m(R_o,\gamma)=\frac{3Z^2\lambda_B}{10R_o}~\frac{2-5\gamma^3+3\gamma^5}{(1-\gamma^3)^2},
\label{Um2}
\end{equation}
where $\gamma\equiv R_i/R_o\leq 1$. With increasing shell thickness (i.e., decreasing $\gamma$),
$U_m$ from Eq.~(\ref{Um2}) {\it increases}, exceeding $U_m$ for a surface-charged microgel 
[Eq.~(\ref{Um})], and approaching the result for a volume-charged microgel as $\gamma\to 0$.
Thus, spreading the fixed charged over the volume of a spherical shell would increase 
electrostatic self-energy and thereby tend to promote swelling.

\begin{figure}[h]
\includegraphics[width=0.9\columnwidth]{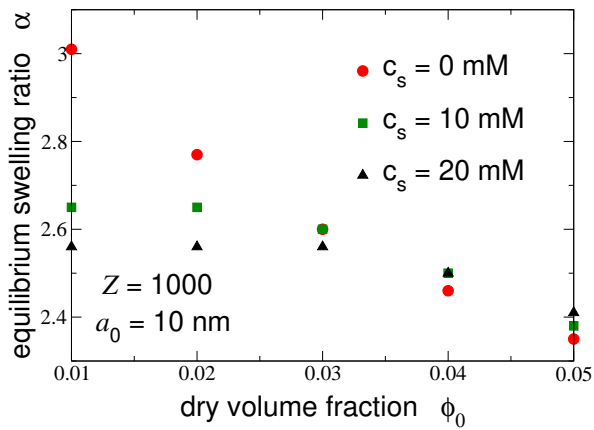}
\vspace*{-0.2cm}
\caption{Equilibrium swelling ratio $\alpha$ vs.~dry volume fraction $\phi_0$ for 
surface-charged microgels of valence $Z =1000$ and dry radius $a_0=10$ nm 
in aqueous suspensions with system salt concentrations $c_s=0$ (red circles), 
10 mM (green squares), and 20 mM (black triangles).
Over this range of $\phi_0$, the swollen volume fraction, $\phi=\alpha^3\phi_0$,
varies (nonlinearly) from about 0.28 to 0.67 for surface-charged microgels and 
the fixed charge concentration $c_f$ varies from about 4 to 20 mM.}
\label{figsalt}
\end{figure}

Finally, we explore the response of ionic microgels to changes in salt concentration. 
For two different system salt concentrations, $c_s=10$ and 20 mM, we numerically solved 
the PB equation [Eq.~(\ref{pb1})] for both counterion and coion density profiles 
$n_{\pm}(r)$, from which we computed the electrostatic component of the single-microgel 
osmotic pressure from Eq.~(\ref{pie2}) and equilibrium swelling ratios from Eq.~(\ref{pim1}).
Figure~\ref{figsalt} shows the resulting variation of equilibrium swelling ratio with 
microgel concentration for the same system parameters as in Fig.~\ref{profile}.
Over this range of dry volume fraction, $0.01<\phi_0<0.05$, the average concentration 
of fixed charge,
\begin{equation}
c_f = \frac{3Z}{4\pi}~\frac{\phi_0}{(a_0 [{\rm nm}])^3}~\frac{1~{\rm mM}}{6.022\times 10^{-4}~{\rm nm}^{-3}}~,
\end{equation}
varies from about 4 to 20 mM. Since $c_f$, and thus also the counterion concentration, 
is comparable in magnitude to the salt concentration, some significant response of swelling 
to variation of $c_s$ may be expected.
At higher salt concentrations, the nonlinear PB equation becomes so numerically stiff that 
our differential equation solver unfortunately fails. However, even for $c_s\leq 20$ mM, 
we already observe a significant influence of added salt on microgel swelling.

In relatively dilute suspensions ($\phi_0<0.03$), microgels typically respond to 
increasing salt concentration by deswelling. This response is conventionally attributed 
to enhanced screening of the microgel fixed charge upon increasing the overall concentration 
of microions and the attendant suppression of electrostatic-driven swelling.
Another interpretation is suggested, however, by close inspection of the expression for 
the electrostatic component of the single-microgel osmotic pressure [Eq.~(\ref{pie2})]. 
Adding salt ion pairs to the suspension leads on average to microgels preferentially 
absorbing counterions over coions (i.e., $\la N_+\ra$ rises more than $\la N_-\ra$). 
Dilute suspensions provide ample free volume outside of the microgels in which coions 
can minimize their energetically costly interactions with microgel surfaces.
With the electrostatic component of the osmotic pressure being thus weakened, 
the gel component favors a lower swelling ratio. 

In more concentrated suspensions ($\phi_0>0.03$), we observe an interesting reversal 
in the swelling response, with crowded microgels swelling slightly upon adding salt.
This unusual response implies that highly crowded microgels, which already harbor a 
significant concentration of counterions, tend to engulf more of any added coions than 
added counterions. This trend is perhaps not surprising, considering that in crowded 
suspensions, where free volume outside of the microgels is limited, exterior coions 
are constrained to be relatively close to the like-charged microgel surfaces, where 
they must pay a higher energetic cost of interaction. In contrast, by moving inside 
a microgel, a coion can avoid the charged surface, while also reducing energy by gaining 
proximity to oppositely-charged counterions. Swelling of microgels further favors coions 
remaining inside the microgels. This result illustrates the complex nature of the 
electrostatic component of the single-microgel osmotic pressure, which involves a 
delicate balance between the mean numbers of counterions and coions inside the microgels.

Whether the swelling trend observed with increasing salt concentration would continue 
to higher microgel concentrations is an interesting question. Unfortunately, our 
cell-model approach is unable to provide a clear answer, since $\phi_0=0.05$ already 
approaches a swollen volume fraction at which elastic interactions between neighboring 
microgels would start to become important. This question could be addressed, however, 
within a multi-center model that includes microgel-microgel interactions~\cite{weyer-denton2018}.
Finally, we note that, despite the qualitatively different swelling behavior of 
ionic microgels as compared with that of bulk ionic gels, our approach has in common with
theories of bulk gels~\cite{tanaka1982,tanaka1984,siegel1988,barrat-joanny-pincus1992}
that it associates swelling with osmotic pressure -- albeit that of a single microgel --
which is intimately related to microion distributions inside and outside of the gel.

\section{Conclusions}\label{conclusions} 
Within the primitive and cell models of permeable, compressible, ionic microgels, 
we investigated the dependence of single-particle osmotic pressure and equilibrium 
swelling on the distribution of fixed charge over the cross-linked polymer network
making up the particles. Applying an exact theorem for the electrostatic component of 
the single-particle osmotic pressure -- implemented within Poisson-Boltzmann theory
and validated via molecular dynamics simulations -- and modeling the gel component
using Flory-Rehner theory of polymer networks, we computed equilibrium swelling 
ratios of microgels (with monovalent counterions) whose fixed charge is confined 
to and uniformly distributed over their surfaces. 

Our study demonstrates that the osmotic pressure and swelling properties of ionic 
microgels are strongly dependent on the fixed charge distribution through its
influence on the microgel electrostatic self-energy and the microion density profiles.
Our results further indicate that surface-charged microgels deswell significantly 
more strongly with increasing concentration (i.e., crowding) than microgels 
that carry a fixed charge uniformly distributed over their volume. Results for these 
two idealized models may serve to bracket the swelling behavior to be expected
of ionic microgels with intermediate fixed charge distributions. Finally, we found 
that, with increasing salt concentration, swelling of surface-charged microgels 
in dilute suspensions is suppressed, due to enhanced screening of bare Coulomb 
electrostatic interactions, but that the effect is weaker, or can even invert, 
in more concentrated suspensions.

Although our study is restricted to suspensions of relatively idealized model 
microgels with uniform cross-link densities and fixed charges evenly distributed 
over their surfaces or volumes, our results vividly reveal the importance of 
fixed charge distribution for single-particle osmotic pressure and equilibrium swelling. 
Further increasing the salt concentration or varying counterion 
and microgel valences may significantly modify electrostatically-driven swelling. 
Steric interactions associated with nonzero microion size also may be important. 
While such effects are beyond the scope of the present mean-field modeling approach,
they would be well worth exploring in future.

Our conclusions regarding the influence of fixed charge distribution
on swelling are qualitatively consistent with experimental measurements of ionic microgels 
and can be further tested in experiments (e.g., light or neutron scattering)
on microgels produced via synthesis protocols that yield varying distributions of fixed charge. 
Thus, our results may provide insights into swelling behavior and guide 
practical approaches to controlling swelling of microgels for practical applications.  
In future work, we will explore structural properties of bulk suspensions of
surface-charged ionic microgels and of mixtures thereof with size and charge asymmetry.

\acknowledgments
We thank Mariano Brito, Gerhard N\"agele, Jan Dhont, and Andrea Scotti for 
helpful discussions. MA acknowledges Shaqra University for financial support.
ARD acknowledges support of the National Science Foundation (Grant No.~DMR-1928073).

\begin{center}
{\bf DATA AVAILABILITY}
\end{center}

The data that support the findings of this study are available from the corresponding author
upon reasonable request.



%

\end{document}